\documentclass[twocolumn,prl,showpacs]{revtex4}

\usepackage{graphicx}   
\usepackage{dcolumn}    
\usepackage{bm}         

\begin{document}

\title{No light shining through a wall : new results from a photoregeneration experiment}

\author{C. Robilliard$^{1}$}


\author{R. Battesti$^2$}
\author{M. Fouch\'e$^1$}
\author{J. Mauchain$^2$}
\author{A.-M. Sautivet$^3$}
\author{F. Amiranoff$^3$}
\author{C. Rizzo$^1$}

\affiliation{ $^{1}$Laboratoire Collisions Agr\'{e}gats
R\'{e}activit\'{e}, UMR 5589 CNRS - Universit\'e Paul Sabatier Toulouse 3, IRSAMC, 31062 Toulouse cedex 9, France.\\
$^2$Laboratoire National des Champs Magn\'etiques Puls\'es, UMR5147
CNRS-INSA-UPS,
143 Avenue de Rangueil, 31400 Toulouse cedex, France.\\
$^3$Laboratoire pour l'Utilisation des Lasers Intenses, \'Ecole Polytechnique, CNRS, CEA, UPMC, 91128 Palaiseau, France.}

\date{\today}

\begin{abstract}
Recently, axion-like particle search has received renewed
interest. In particular, several groups have started ``light
shining through a wall'' experiments based on magnetic field and
laser both continuous, which is very demanding in terms of
detector background. We present here the 2$\sigma$ limits obtained
so far with our novel set-up consisting of a pulsed magnetic field
and a pulsed laser. In particular, we have found that the
axion-like particle two photons inverse coupling constant $M$ is
$> 8\times 10^5$ GeV provided that the particle mass $m_\mathrm{a}
\sim$ 1 meV. Our results definitively invalidate the axion
interpretation of the original PVLAS optical measurements with a
confidence level greater than 99.9\%.
\end{abstract}


\maketitle

The axion was first proposed 30 years ago to solve the strong CP
problem \cite{Peccei1977}, but other models also support the
existence of such light, neutral, spin-zero bosons
\cite{Svrcek2006,DarkMatter} called axion-like particles. Although
no axion has been definitely detected yet, several experiments and
astronomical observations have limited the range of possible
axion-like particle mass $m_\mathrm{a}$ and inverse axion-like
particle two photons inverse coupling $M$
\cite{Raffelt2006review}.

Last year, an italian collaboration (PVLAS) announced an
unexpected observation of a magnetic dichroism in vacuum which
they suggested might be due to photoregeneration of axion-like
particles \cite{PVLAS2006}. However, their mass and two photon
inverse coupling constant inferred from these PVLAS measurements
were seriously inconsistent with the CAST limits \cite{CAST2007},
albeit the latter are model dependent. There was an urgent need
for a direct independent experimental test of the observed
dichroism \cite{Lamoreaux2006}.

All that has raised a renewed interest in axion-like particle
search, in particular for model independent purely
laboratory-based experiments \cite{Jaeckel2007}. The most popular
set-up, commonly called ``light shining through a wall'', is a
photoregeneration experiment based on the Primakoff effect
coupling an axion-like particle with two photons (a real one from
the laser field and a virtual one from an external magnetic field)
\cite{VanBibber1987}. The experiment consists of converting
photons into axion-like particles of identical energy in a
transverse magnetic field, then blocking the photon beam with a
wall. The axion-like particles hardly interact with the wall and
are converted back to photons in a second magnet. Finally, the
regenerated photons are counted with an appropriate detector. Such
an experiment was conducted in the 90's by the BFRT collaboration
without detecting any regenerated photon signal, which led to
limits on the axion parameters \cite{BNL1993}. Mainly motivated by
the PVLAS astonishing results, several "light shining through a
wall" experiments have been proposed and are currently under
construction \cite{patras} ; at DESY the Axion-Like Particle
Search project (ALPS) ; at CERN, the Optical Search for QED vacuum
magnetic birefringence, Axions and photon Regeneration project
(OSQAR) ; at Jefferson Laboratory, the LIght PseudoScalar Search
project (LIPSS) ; at Fermilab, the GammeV Particle Search
Experiment project. Eventually, the PVLAS collaboration disclaimed
their previous observations \cite{Zavattinibis}.

Experimentally, the main difficulty lies in detection. The
expected regeneration rate is indeed very weak -- less than
$10^{-20}$ -- so that optical shielding has to be perfect and the
detector background very low.

In this letter, we detail our project, and we present the limits
on the axion-like particle mass and two photons inverse coupling
constant we have obtained so far. We have found an original and
efficient way to solve the detection problem as both the laser and
the magnetic field are pulsed, as well as our detector. Contrary
to other similar experiments requiring long integration times, we
are not limited by the background of the detector as the photons
are concentrated in very intense and short laser pulses. We are
the first to present here the results of a pulsed "light shining
through a wall" experiment, specially designed to test the PVLAS
claims. In particular, we have found that the axion-like particle
two photons inverse coupling constant $M$ is $> 8\times 10^5$ GeV
provided that the particle mass $m_\mathrm{a} \sim$ 1 meV. Our
results definitively invalidate the axion interpretation of the
original PVLAS optical measurements with a confidence level
greater than 99.9\%.

\begin{figure}
\begin{center}
\resizebox{1\columnwidth}{!}{
\includegraphics{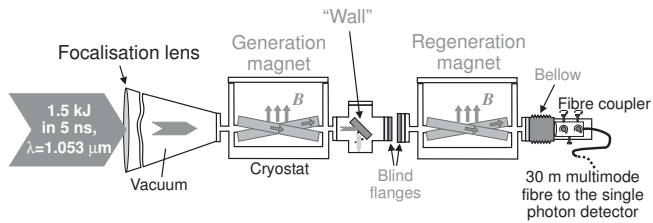}
} \caption{Scheme of our experimental setup.} \label{fig:Setup}
\end{center}
\end{figure}

Our experimental setup shown in Fig.\,\ref{fig:Setup} is based on
three synchronized pulsed elements:  a very energetic laser, two
pulsed magnets which are placed on each side of the wall and a
time-gated single photon detector. We have chosen this pulsed
approach as it allows us to measure very small conversion rates
free from the inevitable background counts of photon detectors.

The conversion and
reconversion transition rate (in natural units $\hbar=c=1$, with 1
T $\equiv 195$ eV$^2$ and 1 m $\equiv 5\times 10^6$ eV$^{-1}$)
after propagating over a distance $z$ in the inhomogeneous
magnetic field $B$ writes \cite{Sikivie}:

\begin{equation}
P\left(z\right) = \left| \int_0^z dz' \Delta_M\left(z'\right) \times
\exp(i \Delta_\mathrm{a} z') \right|^2, \label{eq:P_integral}
\end{equation}
\noindent where $\Delta_M = \frac{B}{2M} \quad \mbox{and}\quad
\Delta_\mathrm{a} = - \frac{m_\mathrm{a}^2}{2\omega}$ with the
photon energy $\omega$. Note that this equation is valid both for
pseudoscalar and scalar particles, but pseudoscalar ({\it resp.}
scalar) particles couple to photons with a polarization parallel
({\it resp.} orthogonal) to the magnetic field. We have two
identical magnets, the detection rate of regenerated photons is
given by
\begin{equation}
R = P^2\frac{\mathcal P}{\omega}{\eta},
\label{eq:regenrate}\end{equation} with $\mathcal P$ the power laser
and $\eta$ the global detection efficiency.

Studying Eqs.\,(\ref{eq:P_integral}) and (\ref{eq:regenrate}), we
can easily see that the number of incident photons, the integral
of the transverse magnetic field over the magnet length $L$:
\begin{equation}
\int_{-L/2}^{+L/2} B dz = B_0 L_{\mathrm{eq}},
\end{equation}
and the detection efficiency have to be maximized. We define $B_0$
as the maximum field and $L_{\mathrm{eq}}$ as the equivalent
length of a magnet producing a uniform magnetic field $B_0$. On
the other hand, $P(z)$ oscillates for too long magnets. The length
leading to the highest conversion rate for a homogeneous magnetic
field is $L_{\mbox{\scriptsize opt}}=2\pi\omega /m_\mathrm a^2$.
For optical frequencies and an axion-like particle mass on the
order of 1 meV, this length is on the order of 1 m.

\begin{figure}
\begin{center}
\resizebox{1\columnwidth}{!}{\includegraphics{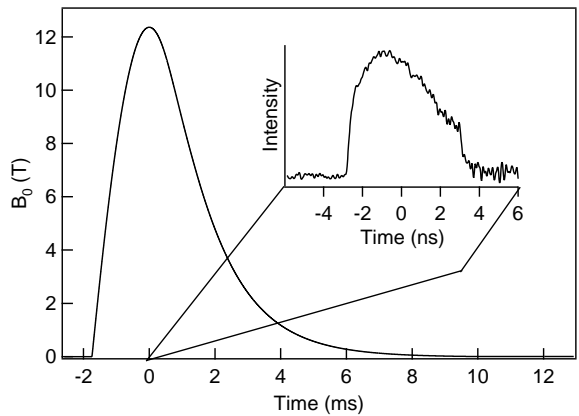}}
\caption{Magnetic field $B_0$ at the center of the magnet as a
function of time. The maximum is reached within 1.75\,ms and can
be considered as constant ($\pm 0.3\%$) during $\tau_{B} =
150\,\mu$s. The 5\,ns laser pulse is applied during this interval.
Inset: Temporal profile of the laser pulse.} \label{fig:B_time}
\end{center}
\end{figure}

In order to have the maximum number of incident photons for the
laser source at a wavelength that can be efficiently detected, we
have chosen to set up the experiment at LULI, Palaiseau, France,
on the Nano 2000 chain. It can deliver up to 1.5\,kJ over 4.8\,ns
(FWHM) --as shown in the inset of Fig.\,\ref{fig:B_time} -- with
$\omega = 1.17$\,eV. This corresponds to $N_{\mathrm{inc}} =
8\times10^{21}$\, photons per pulse. The repetition rate is
1\,pulse every 2\,hours. The vertically linearly polarized
incident beam has a 186\,mm diameter and is almost perfectly
collimated. A deformable mirror included in the middle of the
amplification chain corrects the spatial phase of the beam to
obtain at focus a spot better than two diffraction limits. It is
then focused just behind the wall using a lens which focal length
is 20.4\,m. The beam is apodized to prevent the incoming light
from generating a disturbing plasma on the sides of the vacuum
tubes. Before the wall where the laser beam propagates, a vacuum
better than $10^{-3}$\,mbar is necessary in order to avoid air
ionization. Two turbo pumps along the vacuum line give
$10^{-3}$\,mbar near the lens and better than $10^{-4}$\,mbar
close to the wall. The wall is made of a 15\,mm width aluminum
plate to stop every incident photon while axion-like particles
continue. It is tilted by 45\,$^\circ$ compared to the axis of the
laser propagation in order to increase the area of the laser
impact and to avoid back reflected photons. In the second magnetic
field region, a vacuum better than $10^{-3}$\,mbar is also
maintained.

For the magnets, we use a pulsed technology. The pulsed magnetic
field is produced by a transportable generator developed at LNCMP,
Toulouse, France, which consists of a capacitor bank releasing its
energy in the coils in a few milliseconds \cite{Frings}. A typical
time dependence of the magnetic field in our coils is shown in
Fig.\,\ref{fig:B_time}. Besides, a special coil geometry has been
developed in order to reach the highest and longest transverse
magnetic field \cite{BMVprototype}. A 12\,mm diameter aperture has
been made inside the magnets for the laser beam. As for usual
pulsed magnets, the coils are immersed in a liquid nitrogen
cryostat to limit the consequences of heating. When the magnetic
field is maximum, the repetition rate is set to 5 pulses per hour.
A delay between two pulses is necessary to get back to the
temperature of equilibrium which is monitored via the coil
resistance. During data acquisition, our coils provide $B_0 \geq$
12.3\,T over an equivalent length $L_{\mathrm{eq}} = 365$\,mm. The
magnetic field $B_0$ remains constant ($\pm 0.3\%$) during
$\tau_{B} = 150\,\mu$s, a very long time compared to the 5\,ns
laser pulse. During operation the magnetic pulse is triggered by a
signal from the laser chain which has a stability ensuring that
the laser pulse happens within these 150 $\mu$s. In order to
detect pseudoscalar particles, the transverse magnetic field is
parallel to the laser polarization.

The last principal element is the single photon detector that has
to meet several criteria. In order to have a sensitivity as good
as possible, the regenerated photon detection has to be at the
single photon level. The integration time is limited by the 5\,ns
laser pulse. This imposes a detector with a dark count far lower
than 1 over this integration time so that a non zero regenerated
photon counting would be significant.

Our detector is a commercially available single photon receiver
from Princeton Lightwave which has a high detection efficiency at
$1.05\,\mu$m. It integrates a $80\times80\,\mu$m$^2$ InGaAs
Avalanche Photodiode (APD) thermoelectrically cooled, with all the
necessary bias, control and counting electronics. Light is coupled
to the photodiode through a FC/PC connector and a multimode fiber.
When the detector is triggered, the APD bias voltage is raised
above its reverse breakdown voltage $V_{\mathrm{br}}$ to operate
in ``Geiger mode''. For our experiment, the bias pulse width is
5\,ns to correspond with the laser pulse.

The APD bias voltage is then adjusted to obtain the best
compromise between the detection efficiency and the dark count
rate per pulse. The detection efficiency $\eta$ is measured by
illuminating the detector with a calibrated laser intensity, $\eta
= 0.50 (0.02)$. The dark count rate is about $5\times10^{-4}$
counts per pulse.

After the second magnet, regenerated photons are injected into the
detector through a coupling lens plus a graded index multimode
fiber with a 62.5\,$\mu$m core diameter, a 0.27 numerical aperture
and an attenuation lower than 1\,dB/km. These parameters ensure we
can easily inject light into the fiber with a high coupling ratio,
even when one takes into account the pulse by pulse instability of
the propagation axis that can be up to 9\,$\mu$rad. During data
acquisition, a typical coupling efficiency through the fibre was
found to be about $\eta_c = 0.85$. This efficiency is measured by
removing the wall and the blind flanges (see Fig. 1), and by using
the laser beam from the pilot oscillator without chopping nor
amplifying it. This procedure ensures that the pulsed kJ beam is
perfectly superimposed to the alignment beam.

The only remaining source of misalignment lies in thermal effects
during the high energy laser pulse, which could slightly deviate
the laser beam, hence generating supplementary losses in fibre
coupling. This misalignment is reproducible. This means that it
can be corrected by properly changing the initial laser pointing.
By monitoring the optical path followed by the high energy beam
for each pulse, we were able to take such misalignment losses into
account, and we have observed a maximum value of $20 \%$ of
coupling reduction.

\begin{figure}
  \includegraphics[width=8cm]{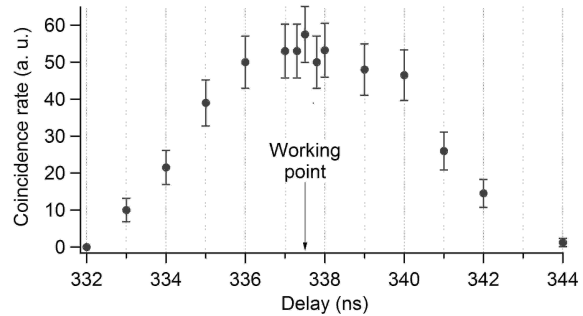}\\
  \caption{Coincidence rate between the arrival of photons on the
detector and its 5 ns detection gate as a function of an arbitrary
delay time. The arrow indicates our working point, chosen in order
to maximize the coincidence rate.}\label{Coinc}
\end{figure}

The detector gate is triggered with the same fast signal as the
laser, using delay lines. We have measured the coincidence rate
between the arrival of photons on the detector and the opening of
the 5 ns detector gate as a function of an adjustable delay. We
have chosen our working point in order to maximize the coincidence
rate (see fig. \ref{Coinc}). To perform such a measurement we used
the laser pilot beam which was maximally attenuated and chopped
with a pulsed duration of 5 ns, exactly as the kJ beam.

The fiber to inject the detector is 30\,m long so that it can be
placed far from the magnets to avoid potential electronic noise
during magnetic shots. In addition the detector is placed in a
shielding bay to prevent electro-magnetic noise during laser
pulses.

So far, during data acquisition, a total amount of about 17.4 kJ
has reached the wall in 14 different pulses. This corresponds to
about 9.3$\times 10^{22}$ photons. To evaluate the actual number
of incident photons that could yield a regenerated photon
observable by the detector, we took into account for each pulse
the fibre coupling $\eta_c$, the misalignment due to thermal
effects during the pulse. We have also evaluated the percentage of
the whole laser energy (see inset of Fig.  2) actually contained
in the 5 ns detection gate, which is 93 $\%$. All these
experimental parameters are known with a few percent errors. The
effective number of photons is about $6.7\times 10^{22}$, which
corresponds to about 12.5 kJ. No regenerated photon has been
detected. In this case, the measurement error is given by the
number of photons that could have been missed due to the non
perfect detection. The probability $P_n$ that $n$ incident photons
have been missed by the detector is $P_n = (1-\eta)^n$. Dark count
is negligible. A standard deviation $\sigma$ means that a result
outside the window $\pm 2\sigma$ corresponds to $P_n < 0.05$,
which yields about 4 missed photons for our value of $\eta$.

\begin{figure}
\begin{center}
\resizebox{1\columnwidth}{!}{
\includegraphics{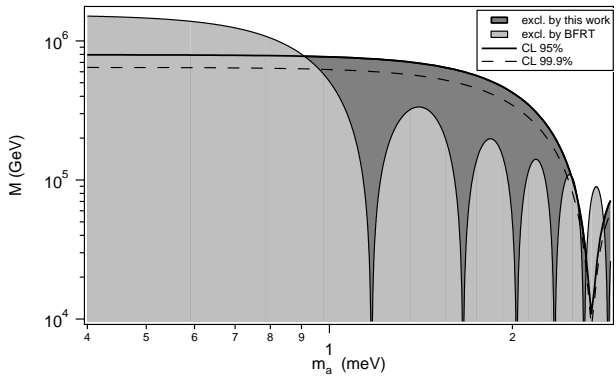}
} \caption{95\% confidence level limits on the axion-like particle
two photons inverse coupling constant $M$ as a function of the
axion-like particle mass $m_\mathrm{a}$ obtained thanks to our
null result  (dotted line). The area below our curve is excluded.
Our limits are compared to the $95\,\%$ confidence level exclusion
region obtained by the BFRT photon regeneration experiment
\cite{BNL1993}.} \label{fig:LULI_Br}
\end{center}
\end{figure}

The limits at $95\,\%$ and 99.9\,\% confidence level that we have
reached so far are plotted on Fig.\,\ref{fig:LULI_Br}. These have
been calculated by numerically solving Eq.\,(\ref{eq:P_integral}).
The area below our curve is excluded by our null result. In
particular the axion-like particle two photons inverse coupling
constant $M$ is $> 8\times 10^5$ GeV provided that the particle
mass $m_\mathrm{a} \sim$ 1 meV. This improves the exclusion region
obtained on BFRT photon regeneration experiment \cite{BNL1993}. In
this mass region their results were limited by the axion-like
particle photon oscillation due to the length of their magnets.
Using shorter magnets we are able to enlarge the mass range
exclusion area.

In ref. \cite{PVLAS2006}, the PVLAS collaboration suggested that
their claimed observation of a vacuum magnetic dichroism could be
explained by the existence of an axion-like particle with a two
photons inverse coupling constant $1\times 10^5 \leq M \leq
6\times 10^5$ GeV and a mass around 1 meV. This is excluded by us
with a confidence level greater than 99.9 \%.

We plan to improve our apparatus so that with about 100 laser
pulses we will be able to give more stringent limits on $M$ than
the one given by the BFRT experiment for all the values of
$m_\mathrm{a}$.\\

We thank the technical staff from LCAR, LNCMP and LULI, especially
S. Batut, E. Baynard,  J.-M. Boudenne, J.-L. Bruneau, D. Castex,
J.-F. Devaud, S. Faure, P. Frings, M. Gianesin, P. Gu\'ehennec, B.
Hirardin, J.-P. Laurent, L. Martin, M. Nardone, J.-L. Paillard, L.
Polizzi, W. Volondat, and A. Zitouni. We also thank B. Girard, G.
Rikken and J. Vigu\'e for strongly supporting this project. This
work has been possible thanks to the ANR-Programme non
th\'{e}matique (Contract ANR - BLAN06-3-139634).

\bibliography{apssamp}

\end{document}